\documentclass[%
 aip,
 amsmath,amssymb,
 reprint,%
]{revtex4-1}

\usepackage{graphicx}
\usepackage{dcolumn}
\usepackage{float}
\usepackage{bm}

\usepackage[utf8]{inputenc}
\usepackage[T1]{fontenc}
\usepackage{mathptmx}
\usepackage{etoolbox}
\usepackage{xcolor}

\makeatletter
\def\@email#1#2{%
 \endgroup
 \patchcmd{\titleblock@produce}
  {\frontmatter@RRAPformat}
  {\frontmatter@RRAPformat{\produce@RRAP{*#1\href{mailto:#2}{#2}}}\frontmatter@RRAPformat}
  {}{}
}%
\makeatother
\begin{document}

\preprint{AIP/123-QED}

\title[Partial vision leads to an unexpected emergent collective behavior in active aligning particles]{Partial vision leads to an unexpected emergent collective behavior in active aligning particles}

\author{Raúl Molina-Prados Lallena}
\affiliation{Departamento de Electrónica, Física Átomica y Térmica, Universidad Complutense de Madrid, Madrid, Spain.}
 \affiliation{GISC, Madrid.}
\author{José Martín-Roca}%
\email{jose.martin.roca@ub.edu}
\affiliation{GISC, Madrid.}
\affiliation{Departament de Física  de la Materia Condensada, Universitat de Barcelona, C. Martı Franques 1, 08028 Barcelona, Spain}

\author{Chantal Valeriani}
\affiliation{Departamento de Electrónica, Física Átomica y Térmica, Universidad Complutense de Madrid, Madrid, Spain.}
\email{cvaleriani@ucm.es}
\affiliation{GISC, Madrid.}

\date{\today}

\begin{abstract}
The Vicsek Model represents a paradigmatic framework for understanding the collective motion of active aligning particles, traditionally assuming isotropic interaction fields. Inspired by biological systems characterized by limited perception and blind spots, we propose a generalized Vicsek model featuring two distinct, non-overlapping angular vision cones. We systematically investigate the non-equilibrium phase behavior of this system by tuning the aperture area ($\alpha$) and the front-back orientation ($\beta$) of the cones. Our results reveal that restricting the lateral vision area destabilizes global order, shifts the critical noise, and induces highly dense traveling bands. Furthermore, breaking the front-back symmetry introduces non-reciprocal interactions that profoundly alter the emergent spatial structures: forward-biased vision drives strong clustering through  "follow the leader" alignment, whereas backward-biased alignment stabilizes an exceptionally homogeneous flocking state with suppressed density fluctuations. Finally, we incorporate short-range volume exclusion, demonstrating that the structural integrity of these novel tightly-clustered phases is highly sensitive to steric interactions. Our work provides new insights into the interplay between non-reciprocal perception, spatial anisotropy, and physical constraints in active matter.
\end{abstract}

\maketitle

\section{Introduction}

Collective motion is a phenomenon observed in a wide range of biological systems, from flocks of birds and fish schools to bacterial colonies and migrating cells \cite{Parrish2002,Hemelrijk2005,Bajec2009}. These systems are paradigmatic examples of active matter, which studies intrinsically out-of-equilibrium systems in which individual agents continuously convert energy into motion: this gives rise to large-scale self-organization without central control \cite{Marchetti2013,bechinger2016,vrugt2025}. Understanding how simple local interactions lead to emergent collective behavior remains one of the central challenges in modern statistical physics.

The Vicsek Model \cite{Vicsek1995} has been recently proposed as  
a minimal model of self-propelled point-like particles  aligning their direction of motion with that of their first neighbors (besides a random angular noise). Despite its simplicity, the model exhibits a rich phase behavior, consisting of a  non-equilibrium phase transition from a disordered state to a collectively moving phase characterized by a long-range polar order \cite{Ginelli2016,Gregoire2004}. 
This transition is accompanied by a diverse phenomenology, including the formation of traveling high-density bands and microphase separation \cite{Chate2008,solon2015phase}, as well as anomalously large density fluctuations in the ordered phase \cite{Chate2024}. The robustness and universality of these features have established the Vicsek model as the prototypical example of active aligning particles and  a milestone in the study of collective motion.

Over the years, numerous variation of the Vicsek Model have been proposed to incorporate more realistic ingredients and explore the different  behaviors observed in active systems. These include the introduction of topological interactions (motivated by experimental observations in bird flocks \cite{Ballerini2008}), steric effects and excluded volume interactions \cite{Murase2026}, several noise implementations and interactions \cite{Jin2025,Jin2025_Net}, and generalizations to multi-species or heterogeneous systems \cite{Lardet2026}. Other studies have explored nematic alignment rules \cite{Patelli2019}, predictive or cognitive interaction mechanisms \cite{Barberis2016,GiraldoBarreto2025} or the behavior of the system in presence of obstacles \cite{martinez2020trapping,martinez2018collective,serna2023influence,Rahmani2021,Codina2022}. Together, these works highlight the sensitivity of collective behavior to the microscopic details of the dynamics, including noise  and  interactions.

A particularly important aspect which has received increasing attention is the role of anisotropy and non-reciprocity in interactions ~\cite{Chen_2017,tang2025reentrant,yang2025characteristics,bagarti2018milling,guo2026emergent,chen2017fore,wang2024condensation}. In contrast to the original Vicsek model, which assumes isotropic interactions within a circular neighborhood, alive units typically possess limited and direction-dependent sensory capabilities \cite{Becco2006}. Experimental and theoretical studies have shown that such anisotropies can significantly affect the stability and structure of collective motion \cite{Solon2022}. More generally, non-reciprocal or fore-aft interactions-in which the effect of particle $i$ onto particle $j$ differs from that of $j$ onto $i$-have been identified as a generic route to unravel novel dynamical states in suspensions of active particles \cite{you2020nonreciprocity,Knezevic2022,Kreienkamp2024_Asym}.

One of the most natural ways to introduce anisotropy is via  vision-cone interactions, according to which particles only interact with neighbors located within a restricted angular sector centered around particle's direction of motion. 
Vision-cone interactions inherently break action-reaction symmetry and provide a minimal framework to study non-reciprocal effects in collective motion. Such models have revealed a wealth of new phenomena, including modified phase transitions, enhanced ordering in narrow cones, and the emergence of directional defect dynamics \cite{Durve2016,Loos2023,Liu2025,Bandini2025,rouzaire2025nonreciprocal,popli2025ordering}.

More recent works have also explored the interplay between anisotropy and pattern formation, showing that directional interactions  lead to complex structures and anisotropic phases \cite{Paul2026_Anisotropic,Negi2024}.

As far as we are aware, the majority of these studies  have considered a simplified angular interactions, typically involving a single symmetric vision cone or a uniform angular restriction. However, perception in alive systems 
is often far more complex, involving multiple sensory regions with different functional roles, such as forward-directed attention, lateral awareness, and blind spots. The impact of such multi-sector perception on collective dynamics remains largely unexplored. 
Although recent work has begun to address non-reciprocal and anisotropic interactions in related models \cite{Du2026,Paul2026_Tracer,Iyer2026}, a systematic investigation of multiple angular vision cones within a Vicsek-like framework is still lacking.

In this work, we introduce a generalized Vicsek Model  incorporating multiple angular vision cones, allowing for the presence of blind spots and asymmetric perception profiles. This modification leads to an interaction matrix which is intrinsically non-symmetric and dynamically dependent on the particles' orientations. We investigate how these features affect the emergence of collective motion, the nature of the ordering transition, and the resulting phenomenology.

\section{Modeling and numerical details}

The traditional two dimensional metric Vicsek Model~\cite{Vicsek1995} consists of $N$ point-like particles moving in two dimensions with constant speed $v_0$. The state of each particle $i$ at time $t$ is defined by its position $\vec{x}_i(t)$ and heading orientation $\theta_i(t) \in [-\pi, \pi]$. To avoid phase-discontinuity issues inherent to scalar angular averages, the alignment dynamics is determined via circular statistics (vector sum). The local unnormalized orientation components of particle $i$ are computed as:
\begin{align}
S_{x,i}(t) &= \sum_{j=1}^{N} n_{ij}(t)\,\cos\theta_j(t), \label{eq:Sx} \\
S_{y,i}(t) &= \sum_{j=1}^{N} n_{ij}(t)\,\sin\theta_j(t), \label{eq:Sy}
\end{align}
where $n_{ij}(t)$ is the neighbor interaction matrix. The average local orientation angle $\bar{\theta}_i(t)$ is extracted using the two-argument arctangent function
\begin{equation}
\left\langle \theta_i \right\rangle(t) = \operatorname{atan2}\Big(S_{y,i}(t),\, S_{x,i}(t)\Big).
\end{equation}
The orientation update for the time step $\Delta t$ is obtained by adding an additive noise term and wrapping the resulting angle back to the $[-\pi, \pi]$ interval:
\begin{equation}\label{theta_update}
\theta_i(t+\Delta t) =\left\langle \theta_i \right\rangle (t)+ \eta\,\xi_i(t),
\end{equation}
where $\eta \in [0,1]$ controls the noise amplitude and $\xi_i(t)$ is a white noise variable uniformly distributed in $[-\pi, \pi]$.  Once orientations are updated, particle positions evolve according to:
\begin{equation}\label{motion}
\vec{x}_i(t+\Delta t) = \vec{x}_i(t) + v_0\,\vec{s}_i(t+\Delta t)\,\Delta t,
\end{equation}
where $\vec{s}_i(t) = (\cos\theta_i(t),\, \sin\theta_i(t))$ represents the updated velocity direction, subject to periodic boundary conditions in a square box of linear size $L$. In the standard metric Vicsek model, the neighbor matrix elements $n_{ij}^\text{V}(t)$ for $j \neq i$ are defined purely by a distance cutoff $R_0$:
\begin{equation}\label{eq:n_v}
n_{ij}^\text{V}(t) = \Theta\left( R_0 - |\vec{x}_j(t) - \vec{x}_i(t)| \right),
\end{equation}
where $\Theta$ is the Heaviside step function, establishing a isotropic circular interaction area of radius $R_0$.  In this case, the neighbor matrix is symmetric, i.e. the neighbor graph is undirected, which means that the connections between neighbors are reciprocal.

\begin{figure}[h!]
    \centering
    \includegraphics[width=0.7\linewidth]{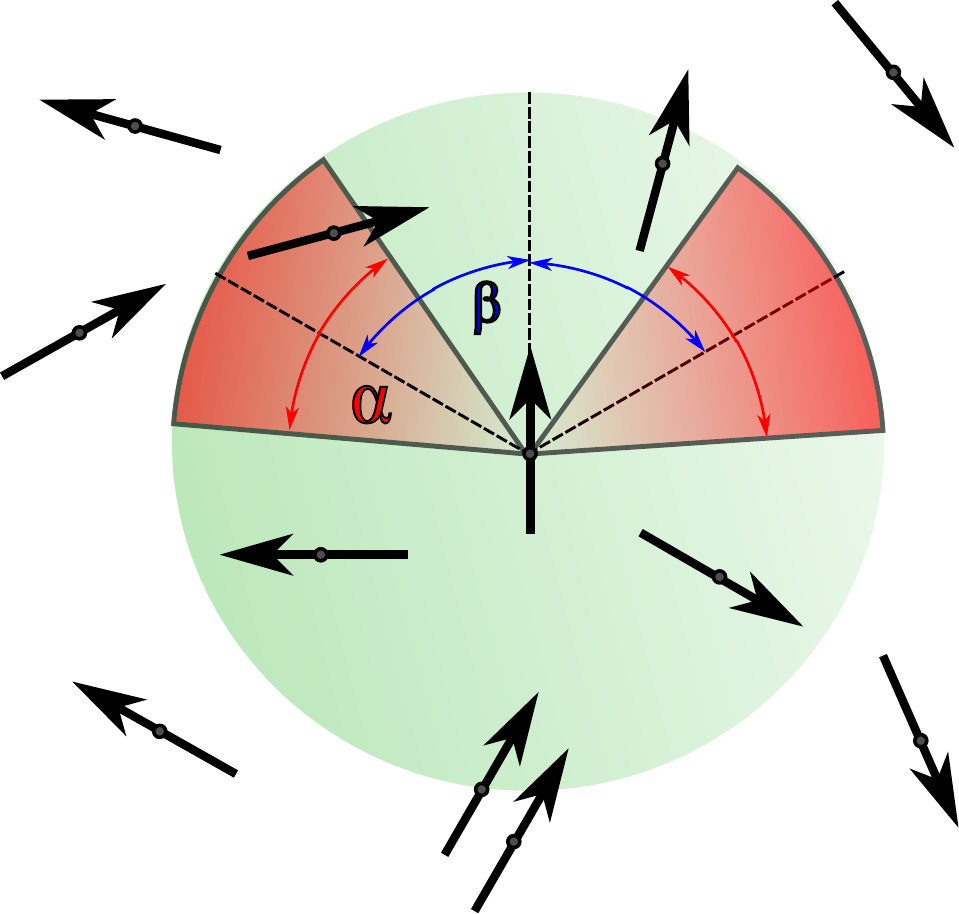}
    \caption{Representation of first neighbors matrix, equation (\ref{eq:n_v}). The arrow in the center represents the agent $i$, while the arrows inside the green region are its first neighbors in Vicsek model and the one inside the red areas (vision cones) are its first neighbors in the modified blind spot Vicsek model.}
    \label{fig:visionCone}
\end{figure}

\subsection{The blind spots Vicsek Model}

The novelty of our proposed blind spot Vicsek model  is to modify it in order to mimic active aligning particles with multiple angular vision cones. As represented in 
Fig.\ref{fig:visionCone}, particles in the proposed blind spot Vicsek Model detect neighbours according to   two cones  located at an angle $\beta$ with respect to  particles direction of motion.  Each vision cone is characterised by an angular width of $\alpha$ (in red).  
Therefore, the active aligning particle at the center of the circle  (Fig.\ref{fig:visionCone}) can only detect as neighbors those particles within the two vision cones (red areas).

{ This fore-aft symmetry breaking leads to a neighbors matrix which is not necessary symmetric (as in the Vicsek case),  i.e. the graph of neighbors is directed, which means that the neighbor connections are non-reciprocal.}  To compute this new matrix, we have been inspired by Nature. As an example,  fish do not have a perfect $360^\circ$ vision, but their vision seems to be characterised by blind spots. This motivated us to estimate particles'  first neighbors matrix as:
\begin{equation}
n_{ij}(t)=
\begin{cases}
1, \quad  & \left[ \; \; \begin{array}{c}
      |\vec{x}_i(t)-\vec{x}_j(t)| \le R_0 \; \text{ and } \; 
 |\theta_i(t)-\theta_j(t)| \in I  \\
 \text{or} \\ 
      |\vec{x}_i(t)-\vec{x}_j(t)| \le R_0 \; \text{ and } \;  |\theta_i(t)+\theta_j(t)| \in I 
\end{array} \right. \\
\; & \; \\
0, & \text{other case},
\end{cases}
\end{equation}
in which we define the interval $I = \left[\,|\beta-\alpha/2|,\;|\beta+\alpha/2|\,\right]$, and $\alpha$ and $\beta$ are graphically reported in the Fig. \ref{fig:visionCone}.  
Note that one recovers the Vicsek model~\cite{Vicsek1995}  for $\alpha=180^\circ$ and $\beta=90^\circ$ { while pure half-from vision~\cite{Chen_2017} is reproduced for $\alpha=90^\circ$ and $\beta=45^\circ$.}

\subsection{Simulation details}
Simulations were performed using an in-house CUDA-based implementation to enable the study of large-scale systems.
Considering as unit length the particles' vision cutoff radius ($R_0=1$), we prepare a two dimensional system in a square domain of size $L=256 R_0^{2}$ with periodic boundary conditions and particle's mean number density $\langle \rho \rangle = \frac{N}{L^2}$, being N the number of particles.
Throughout this work, we set the mean number density as $\langle \rho \rangle = 3.0 R_0^{-2}$; therefore, each simulation contains a total of $N=196,608$ particles. The self-propulsion velocity is set to $v_0=0.5 \, R_0/\tau_0$, being $\tau_0=1.0$ the time unit. 
Simulations were run with a timestep set to $\Delta t=0.1 \, \tau_0/\text{step}$. To ensure that the system successfully reached a true non-equilibrium steady state and to minimize finite-time and critical slowing down effects, the total run-time was adjusted between $10^5$ steps (for states deep within the ordered or disordered phases) and $10^6$ steps (near the phase transition and for extremely narrow vision cones, $\alpha \le 45^\circ$). 
The initial configuration was prepared by randomizing particles' positions and orientations. 
Consequently, the main parameters varied in this study are: (a) the noise amplitude, $\eta$; (b) the vision cone aperture, $\alpha$; and (c) the vision cone orientation (cross-eye angle), $\beta$. In the first part of this work, we study the case of constant lateral vision ($\beta=90^\circ$) while varying $\alpha$ and $\eta$. In the second part, we fix the aperture ($\alpha=90^\circ$) and sweep $\beta$.\\

Finally, to evaluate the role of steric interactions---which are highly relevant in physical experiments such as robotic swarms or active granular matter---we study selected cases in the presence of short-range volume exclusion. Volume exclusion is modeled via a soft-core repulsive potential that modifies the particles' update rule. Since we are particularly interested in small-scale experimental setups with a limited number of agents, we perform these specific simulations for reduced system sizes using same number density $\langle \rho \rangle = 3.0 R_0^{-2}$ with a system size of $L=8R_0^{2}$. 

In order to incorporate physical collisions while strictly maintaining the classic Vicsek constant-speed constraint, we propose the following modification to the velocity update:
\begin{equation}\label{velocityF}
    \vec{v}\;'(t) = v_0 \frac{\vec{v}_i(t)}{||\vec{v}_i(t)||} = v_0 \frac{ v_0 \, \vec{s}_i(t) + \vec{F}_{i} }{||v_0 \, \vec{s}_i(t) + \vec{F}_{i}||},
\end{equation}
where $\vec{F}_{i}= - \nabla U_i$, with $U_i=\sum_{j \neq i} U_{ij}$ being the total steric potential. The pairwise interaction is given by the standard Weeks-Chandler-Andersen (WCA) repulsive potential:
\begin{equation}\label{WCA}
    U_{ij} = 4 \epsilon \left[ \left( \frac{\sigma}{r_{ij}} \right)^{12}- \left( \frac{\sigma}{r_{ij}} \right)^{6} \right] + \epsilon, \quad \text{for } r_{ij} < 2^{1/6} \sigma,
\end{equation}
and $U_{ij} = 0$ otherwise. This formulation guarantees that the velocity modulus remains strictly constant and equal to $v_0$ for all particles. Consequently, the repulsive interaction only induces a deviation in their orientation, simulating physical boundaries that depend on the particle diameter $\sigma$.

\begin{figure*}[ht!]
    \centering
    \includegraphics[width=0.95\linewidth]{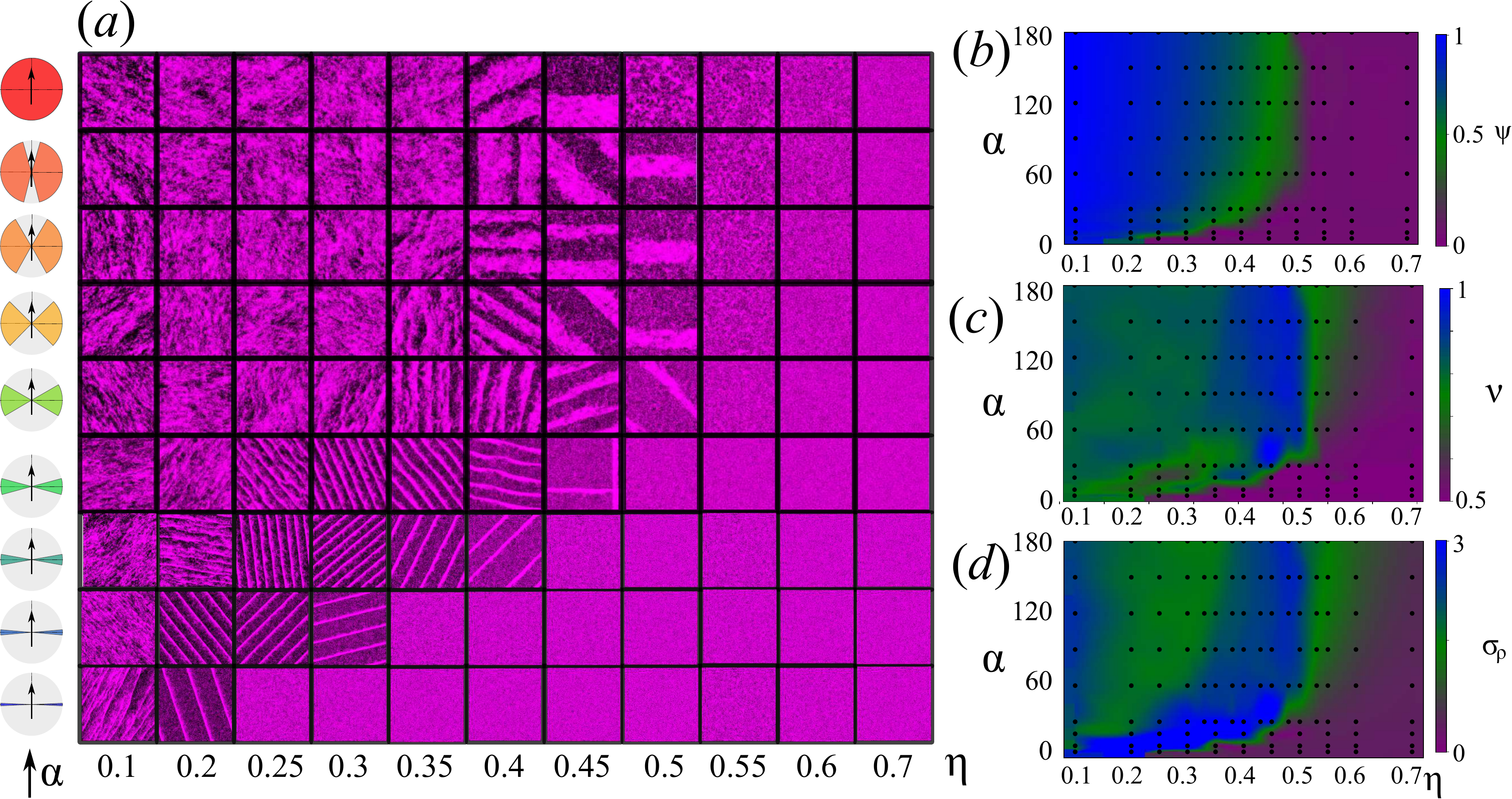}
    \caption{Plane $\alpha$-$\eta$ for $\beta=90^\circ$. (a) Snapshot of the system for different noises and $\alpha \in \{ 5^\circ,10^\circ, 15^\circ,30^\circ,60^\circ,90^\circ, 120^\circ, 150^\circ, 180^\circ \}$. Observables to characterize the phases in the system: (b) polar order parameter, (c) giant density fluctuations exponent and (d) average density fluctuations.}
    \label{fig:plane}
\end{figure*}

\section{Detecting order and phase transitions}

In the traditional metric Vicsek model (Eq. \ref{eq:n_v}), the emergence of collective motion is quantified via the global polar order parameter
\begin{equation}
\label{eq:pod}
\vec{\psi}(t)=\frac{1}{N}\sum_{i=1}^{N}\vec{s}_i(t)
\end{equation}
The modulus $\psi(t)=|\vec{\psi}(t)|$ measures the degree of global alignment:  $\psi \approx 1$ corresponds to an ordered phase characterized by a coherent collective motion, while $\psi \approx 0$ indicates a disordered state with no macroscopic alignment. To further characterize the nature of the transition, estimating the 
critical noise at which the transition takes place, we compute the Binder cumulant
\begin{equation}\label{BC}
G(\eta)= 1 -\frac{\langle \psi^4\rangle_t}{3\langle \psi^2 \rangle_t^2}.
\end{equation}
for every value of the noise $\eta$.
This dimensionless quantity provides information on the fluctuations of the order parameter and allows  to distinguish between continuous and discontinuous phase transitions. In particular, a pronounced minimum in $G(\eta)$ near the critical noise $\eta_c$ value is typically associated with a first-order transition, as observed in the traditional Vicsek Model.

In addition to global ordering, density fluctuations play an important role in  characterizing  active matter systems. To quantify them, we define a coarse-grained density field by partitioning the system into square mesh grids  $a_0$,
\begin{equation}
\rho(\mathbf{r})=\frac{N_{\text{cell}}(\mathbf{r})}{a_0^2},
\end{equation}
where $N_{\text{cell}}(\mathbf{r})$ is the number of particles within a given cell. 
The variance of density fluctuations is then given by
\begin{equation}
\sigma_\rho = \sqrt {\langle \rho^2 \rangle - \langle \rho \rangle^2},
\end{equation}
or in terms of particle number fluctuations,
\begin{equation}
\Delta N^2 = \langle N^2 \rangle - \langle N \rangle^2.
\end{equation}
In equilibrium systems, fluctuations obey the central limit theorem, leading to $\Delta N \sim \sqrt{N}$. In contrast, active matter systems such as the Vicsek model exhibit anomalous scaling of the form $\Delta N \sim N^{\nu}$ with $\nu > 1/2$, a hallmark of so-called Giant Density Fluctuations (GDF).\cite{toner2018} These fluctuations reflect the presence of long-range correlations and are closely linked to the broken symmetry of the ordered phase.
To estimate the presence of anomalous density fluctuations in the system, we have considered several  varying values of $a_0$.

An additional relevant quantity in estimated in the present model is the effective local density surrounding  each particle, which depends on the angular structure of the interaction region: the vision density, 
\begin{equation}\label{density}
\rho_{\text{v}} = \frac{\langle N_{\text{v}} \rangle}{ 2 \, \alpha \, R_0^2},
\end{equation}

where $\langle N_{\text{v}} \rangle$ is the average number of interacting neighbors within the vision domain (red region in Fig. \ref{fig:visionCone}). Unlike the standard Vicsek Model, where the interaction region is isotropic, the introduction of angular vision cones makes $\rho_{\text{v}}$ dependent on both the aperture $\alpha$ and the orientation $\beta$ of the cones.

\section{Results}

In this section, we present the collective behavior, emergent spatial structures, and phase transitions of the proposed blind-spot Vicsek Model. To systematically disentangle the physical effects of the different parameters governing the vision field, we divide our analysis into three distinct parts. In Section \ref{sec:alpha}, we explore the impact of restricting the vision area by varying the aperture angle $\alpha$ while keeping a symmetric lateral orientation. In Section \ref{sec:beta}, we investigate the role of non-reciprocal interactions by varying the cross-eye angle $\beta$ at a constant aperture, thereby breaking the front-back symmetry of the particles' perception. Finally, in Section \ref{sec:exclusion}, we introduce steric interactions to evaluate the robustness of the newly identified active phases under volume exclusion, bringing our model closer to realistic experimental scenarios with finite-sized agents.

\begin{figure*}[ht!]
    \centering
    \includegraphics[width=0.95\linewidth]{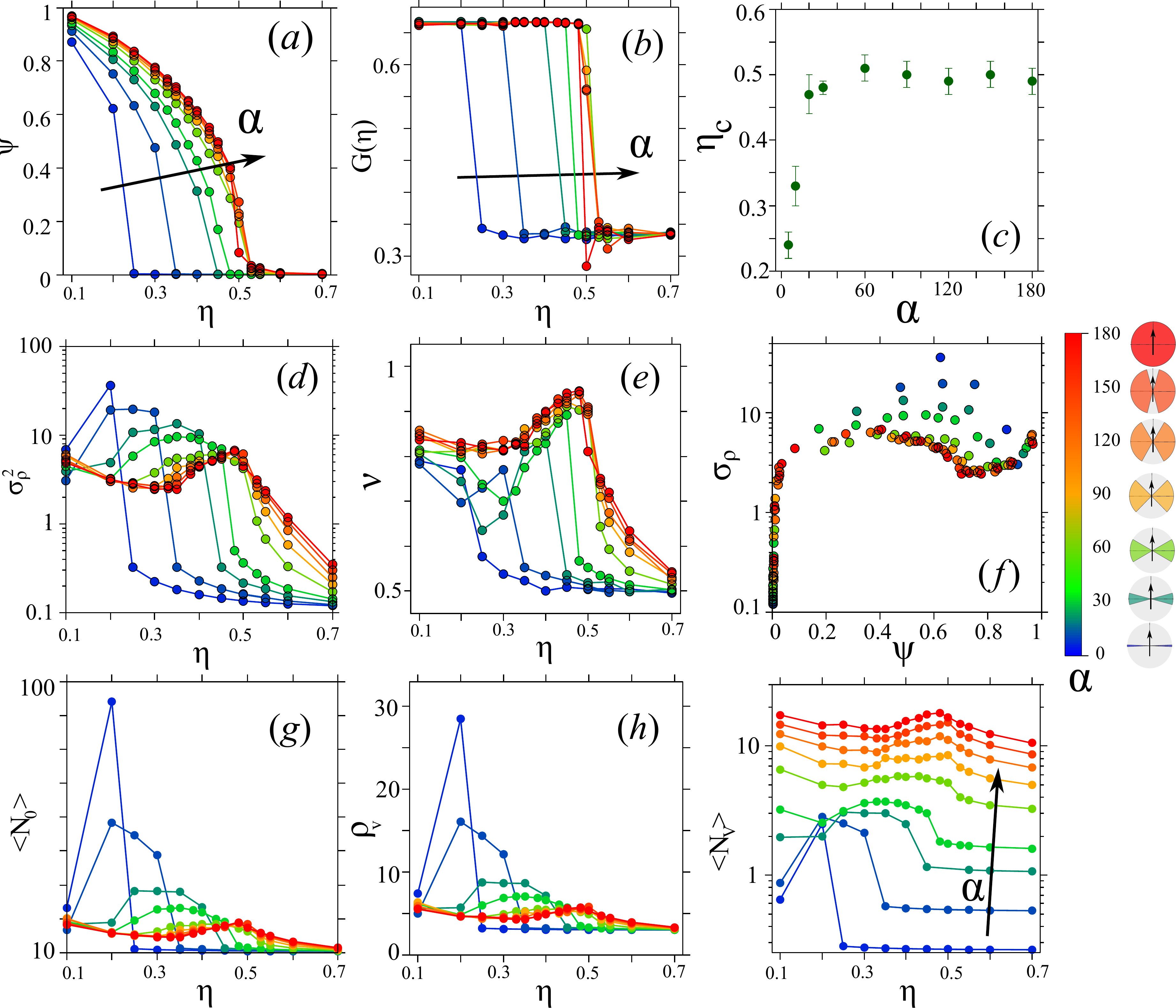}
    \caption{(a) Polar order parameter, $\Psi$; (b) Binder cumulant for the polar order parameter; (c) critical noise, $\eta_c$; (d) density standard deviation, $\sigma_\rho^2$; (e) number fluctuations exponent, $\nu$; (f) polar order vs density fluctuations; (g) number of neighbors at distance $r<R_0$, $\langle N_0 \rangle$;  (h) density of neighbors on vision cones, $\rho_v=N_v/(\alpha \, R_0^2)$, and (i) number of neighbors inside the vision cone, $N_v$, as function of the noise, $\eta$.  All plots are done for different values of $\alpha$ at constant  $\beta=90^\circ$.}
    \label{fig:parameters-alpha}
\end{figure*}

\subsection{Varying the vision area, $\alpha$}
\label{sec:alpha}

We begin by analyzing the effect of the angular aperture $\alpha$ on the collective behavior of the system, fixing the orientation of the cones to $\beta = 90^\circ$. In this configuration, particles interact preferentially with neighbors located laterally with respect to their direction of motion, while forward and backward directions become partially or totally blind depending on $\alpha$. By decreasing $\alpha$, the interaction region is progressively reduced from the isotropic Vicsek case ($\alpha=180^\circ$) to narrow lateral cones.

Figure \ref{fig:plane} summarizes the global behavior of the system in the $\alpha$-$\eta$ plane. As expected, for large values of $\alpha$ the system reproduces the phenomenology of the standard Vicsek Model, exhibiting a transition from a disordered phase at large noise to an ordered phase at low noise. This transition is clearly visible in Fig. \ref{fig:plane}(b), where the polar order parameter $\psi$ sharply decreases as $\eta$ increases.

Figure \ref{fig:plane}(a) shows representative snapshots of the system where the black regions represent void space and the pink arrows indicate individual self-propelled agents. The first row of this figure corresponds to the isotropic Vicsek case, displaying the classical sequence of active phases. At low noise, we find a homogeneous, globally aligned flock, commonly referred to in the literature as the ordered Toner-Tu liquid \cite{Vicsek1995}. Conversely, at high noise, a completely disordered gas phase is established, where particles orient randomly. Near the transition between these two phases, the system exhibits coexistence characterized by the formation of a dense, globally ordered traveling band propagating through a disordered background \cite{solon2015phase, rouzaire2026dynamics,Chen_2017}. This sequence of phases-disordered gas, coexistence with traveling bands, and homogeneous flocking-is consistent with the well-known liquid-gas-like transition scenario of the Vicsek Model driven by self-propelled microphase separation \cite{solon2015phase}. { It should be note that in the cases under study varying $\alpha$ the previously reported highly dense parallel bands~\cite{Chen_2017,tang2025reentrant,yang2025characteristics,bagarti2018milling,guo2026emergent,chen2017fore,wang2024condensation} is not observed, probably due to the absent of back-front symmetry with some front bias, as will be point in the next section.}

By reducing $\alpha$, we observe a similar transition sequence, but with remarkable differences in the morphology of the emerging structures. The traveling bands become narrower and sharper as the vision aperture is restricted. Most notably, for $\alpha=30^\circ$ and $\eta=0.45$, the system exhibits a stable phase characterized by two perpendicular sets of traveling bands forming a grid-like pattern. This distinctive "cross sea" phase has been previously { observe in the standard Vicsek Model for some specific noise values and initial configuration} \cite{kursten2020dry}, yet here it emerges from purely lateral, narrow vision cones.

As the vision aperture $\alpha$ is reduced, several important effects emerge. First, the ordered phase becomes progressively less stable, and the critical noise $\eta_c$ shifts towards lower values, as depicted in Fig. \ref{fig:parameters-alpha}(a) and (c). This values of $\eta_c$ are extracted from the sharp drop of the Binder cumulant $G(\eta)$ (Fig. \ref{fig:parameters-alpha}(b)). Restricting the angular interaction hinders the ability of the system to sustain global alignment. This is a direct consequence of dilution in the interaction network: as $\alpha$ decreases, the effective interaction area $2\alpha R_0^2$ shrinks, leading to a substantial drop in the average number of interacting neighbors within the vision cones, $N_v$ (as confirmed in Fig. \ref{fig:parameters-alpha}(i)). Consequently, particles average their orientations over a smaller local pool of neighbors, making the alignment process highly sensitive to stochastic fluctuations; hence, a much lower noise $\eta$ is sufficient to destroy global order.

The effect of angular restriction is also clearly reflected in density fluctuations. Figure \ref{fig:parameters-alpha}(d) shows the behavior of the density variance $\sigma_\rho^2$, which exhibits a pronounced peak near the transition region. For the disordered phase, density fluctuations are almost zero, whereas they reach their maximum within the coexistence phase (Fig. \ref{fig:parameters-alpha}(d)). Interestingly, the peak value of $\sigma_\rho^2$ is significantly higher for smaller values of $\alpha$. This behavior reveals that restricting the vision fields to lateral cones enhances local clustering at the transition, producing extremely dense and narrow traveling bands that coexist with a highly depleted gas phase, thereby maximizing the density variance.

\begin{figure*}[ht!]
    \centering
    \includegraphics[width=0.95\linewidth]{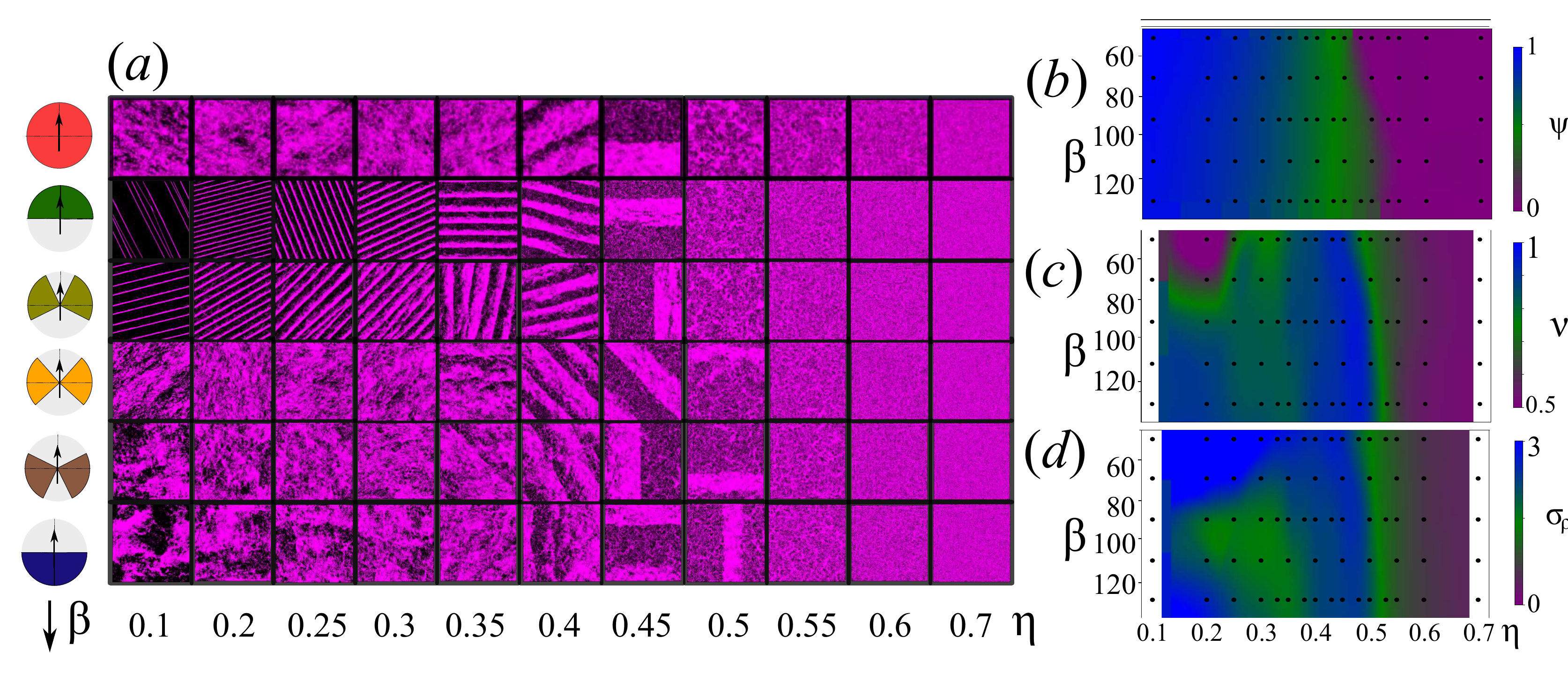}
    \caption{Plane $\beta$-$\eta$ for $\alpha=90^\circ$. (a) Snapshot of the system for different noises, including the isotropic case and $\beta \in \{ 45^\circ,66.5^\circ, 90^\circ,112.5^\circ,135^\circ\}$. (b) polar order parameter, (c) giant density fluctuations exponent and (d) average density fluctuations.}
    \label{fig:plane_beta}
\end{figure*}

To reconcile this with the behavior of the giant density fluctuations (GDF) in the ordered phase, we analyze the scaling exponent $\nu$ (where $\Delta N \sim N^\nu$) in Fig. \ref{fig:plane}(c) and Fig. \ref{fig:parameters-alpha}(e). In the disordered phase, $\nu \approx 0.5$, complying with the central limit theorem. In the ordered phase, the standard Vicsek Model ($\alpha=180^\circ$) yields $\nu > 1/2$, a hallmark of active polar liquids. However, reducing $\alpha$ suppresses these long-range fluctuations, pushing $\nu$ closer to $0.5$ over a wider range of noise. This is because the narrow lateral vision cones break the isotropy of the alignment propagation, preventing the long-range orientation correlations required to sustain large-scale coherent density fluctuations deep within the ordered phase. Therefore, while a smaller $\alpha$ promotes sharper local density segregation at the transition (higher $\sigma_\rho^2$), it simultaneously weakens long-range collective correlations in the homogeneous flocking phase (lower $\nu$).

To further clarify the microscopic structure, we evaluate the local properties shown in Figs. \ref{fig:parameters-alpha}(g)-(i). While the number of interacting neighbors $N_v$ decays with $\alpha$, both the total number of neighbors at distance $r<R_0$ ($\langle N_0 \rangle$) and the effective vision density ($\rho_v$) reveal that, in the ordered-liquid phase (low noise), particles are more tightly packed for smaller aperture angles. This suggests that lateral-only vision forces particles to cluster more densely to maintain alignment. In the disordered phase (high noise), these differences vanish, and the local structure is governed purely by the randomizing noise.

Finally, Fig. \ref{fig:parameters-alpha}(f) illustrates the parametric relationship between the polar order parameter $\Psi$ and the density variance $\sigma_\rho^2$, {as used in previous works to identify traveling bands \cite{chepizhko2021revisiting,knevzevic2022collective}}. For large $\alpha$, we observe a smooth trajectory where density fluctuations peak at intermediate order. As $\alpha$ decreases, the peak of $\sigma_\rho^2$ becomes dramatically sharper and shifts to lower values of $\Psi$. This quantitative shift confirms that the isotropic-to-polar transition changes from a weakly first-order scenario in the standard Vicsek Model to a highly discontinuous transition characterized by extreme spatial clustering and sharp phase separation under restricted lateral vision.

\subsection{Varying the cross-eye degree, $\beta$}
\label{sec:beta}

In this second analysis, we maintain the value of the aperture angle $\alpha=90^\circ$ while varying the orientation of the two vision cones via the cross-eye angle $\beta$ \cite{kursten2020dry}. The main objective here is to understand how the loss of reciprocity and front-back symmetry in interactions affects both the emergent collective patterns and the nature of the phase transition. 

To maintain the physical definition of the model—consisting of two distinct, non-overlapping vision cones separated by front and rear blind spots—the cross-eye angle $\beta$ must be restricted to a specific geometric range. Since each cone has an angular width of $\alpha = 90^\circ$ and is centered at $\pm\beta$ relative to the direction of motion, avoiding overlap at the front requires $\beta - \alpha/2 \ge 0^\circ$. Similarly, preventing the cones from overlapping at the back requires $\beta + \alpha/2 \le 180^\circ$. Consequently, the non-overlapping dual-cone regime restricts the orientation angle to $45^\circ \le \beta \le 135^\circ.$ Outside of this interval, the two cones merge either in the forward or backward direction, violating the distinct dual-perception assumption of our model.

\begin{figure*}[ht!]
    \centering
    \includegraphics[width=0.95\linewidth]{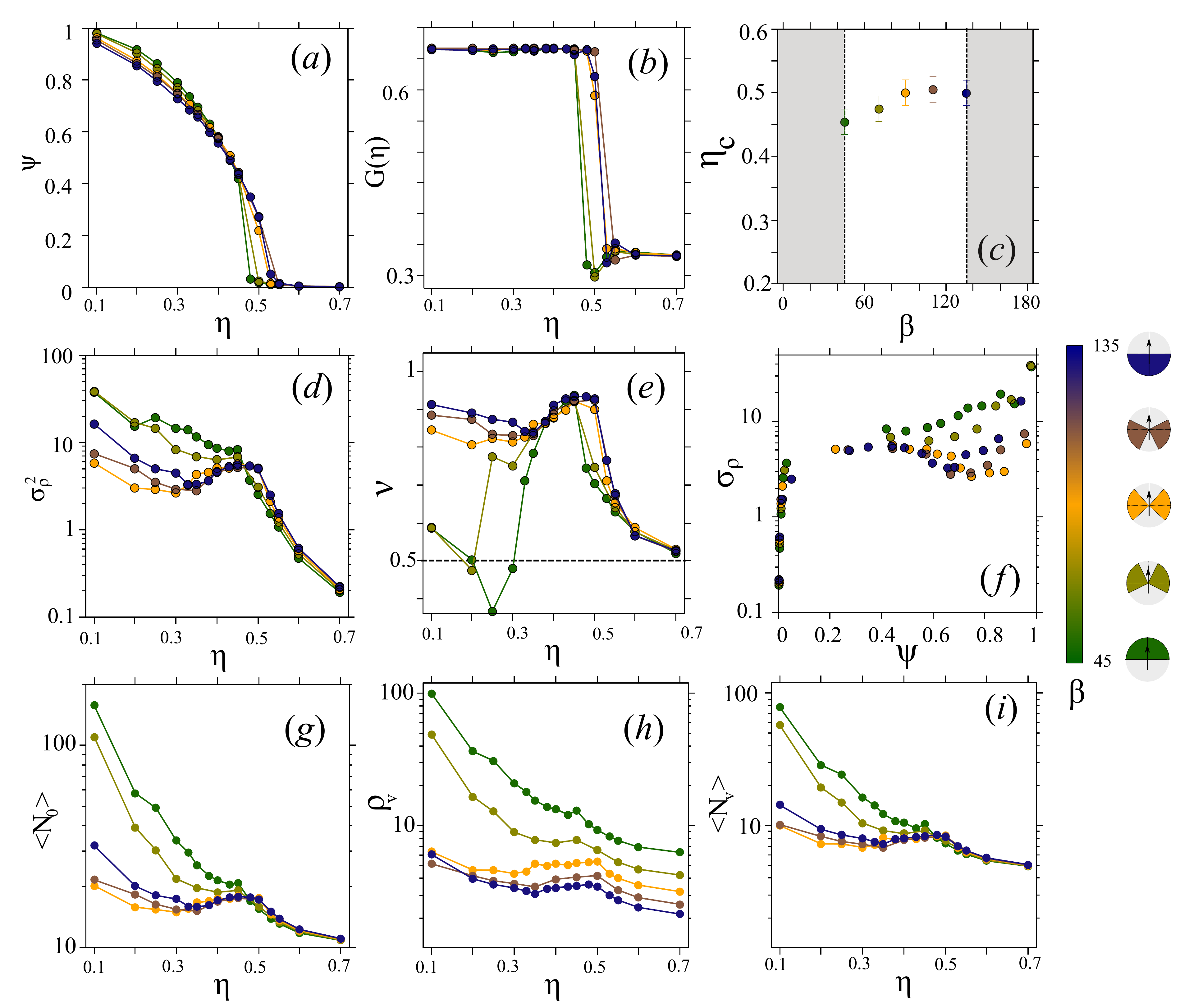}
    \caption{{(a) Polar order parameter, $\Psi$; (b) Binder cumulant for the polar order parameter; (c) critical noise, $\eta_c$, gray region represent the angles for which the two vision cones overlap for the $\alpha$ value under study; (d) density standard deviation, $\sigma_\rho^2$; (e) number fluctuations exponent, $\nu$; (f) polar order vs density fluctuations; (g) number of neighbors at distance $r<R_0$, $\langle N_0 \rangle$; (h) density of neighbors on vision cones, $\rho_v=N_v/(\alpha \, R_0^2)$, and (i) number of neighbors inside the vision cone, $N_v$, as function of the noise, $\eta$. All plots are done for different values of $\beta$ at constant $\alpha=90^\circ$}.}
    \label{fig:parameters_beta}
\end{figure*}

Taking a quick look at the system's snapshots in Fig. \ref{fig:plane_beta}(a), the effects of shifting the vision cones within this non-overlapping regime become very notable. The breaking of front–back symmetry in the interaction field leads to profound modifications in the emergent spatial structures. When the symmetry is broken in favor of the forward direction (i.e., for values of $\beta$ approaching the lower limit of $45^\circ$), the homogeneous ordered-liquid phase completely disappears. Instead, for low noise levels, we find { the well known phase composed of highly dense parallel bands. In contrast to previous works~\cite{Chen_2017,tang2025reentrant,yang2025characteristics,bagarti2018milling,guo2026emergent,chen2017fore,wang2024condensation}, we show that the highly dense band can appears in the system even if the front vision spot is not included in the aliment region provided some bias in the front direction}. This behavior can be understood as a consequence of non-reciprocal chase \cite{you2020nonreciprocity}: when particles preferentially align with those ahead of them, a "follow-the-leader" effect takes place. This feedback mechanism destabilizes homogeneous states and promotes severe spatial clustering into traveling waves \cite{solon2015phase}.\\

Conversely, when the symmetry is broken in favor of the backward direction (i.e., as $\beta$ approaches the upper limit of $135^\circ$), the low-noise phase transitions into an relatively homogeneous ordered state, indicating a strong suppression of spatial clustering compared with the front case. In this backward-biased regime, particles align with those trailing behind them. This acts as a negative feedback loop where leaders adjust their directions to match their followers, which actively prevents the formation of density bands but makes the system less homogeneous than the symmetrical case.  This behavior is clearly reflected in the average density fluctuations shown in Fig. \ref{fig:plane_beta}(d) and the density standard deviation $\sigma_\rho^2$ in Fig. \ref{fig:parameters_beta}(d), both of which are strongly minimized for $\beta = 90^\circ$ and significantly bigger for the forward cases for the low-noise levels.  

This different in terms of density inhomogeneities is also reflected in the behavior of the number fluctuations exponent $\nu$, shown in Fig.\ref{fig:parameters_beta}(e). For the smallest studiend value of $\beta$ within the non-overlapping regime ($\beta=45^\circ$), the exponent appears to drop slightly below the standard equilibrium limit of $0.5$ at moderate noise levels. Although a value of $\nu < 0.5$ is a typical signature of active hyperuniformity (indicating a suppression of density fluctuations beyond the Poissonian limit), we must exercise caution here: this drop is close to the limit and is only observed for specific parameter values. Therefore, rather than claiming the definitive emergence of a hyperuniform phase, we treat this as a tentative and intriguing indicator. A more systematic finite-size scaling analysis would be required to determine whether this anomalous density suppression represents a robust thermodynamic phase or a marginal finite-size crossover effect.

\begin{figure*}[ht!]
    \centering
    \includegraphics[width=0.95\linewidth]{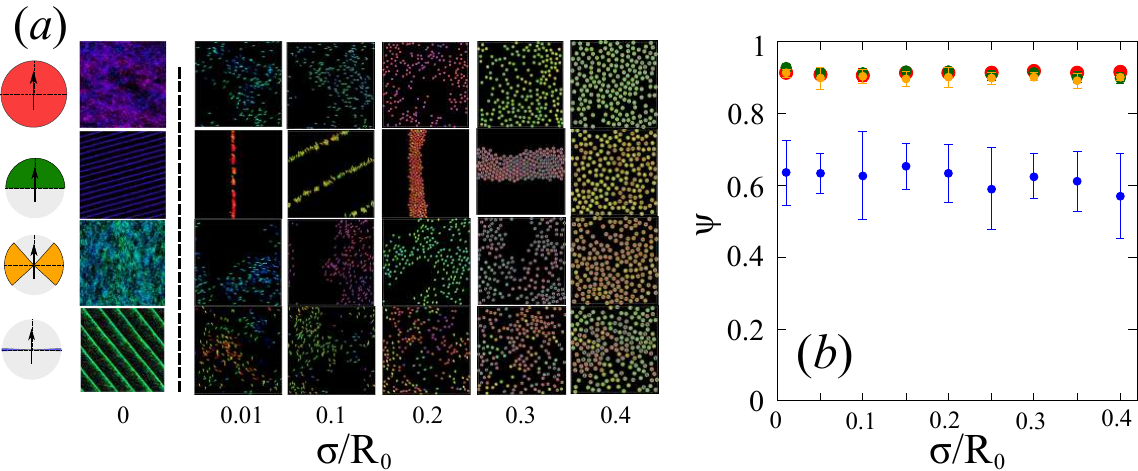}
    \caption{(a) Snapshot of the system for different combinations of $(\alpha,\beta)$ and varying particle size ratios respect to the interaction radius, $\sigma/R_0$. The case for $\sigma/R_0=0$ have been perform in a big system of $N=196.608$ particles as in the other section, while $\sigma/R_0>0$ is for a small system of $N=192$ particles. The arrows are colored using the value of its orientation following the chromatic circle showed at the top left. (b) Polar order parameter $\Psi$ for the different cases illustrated in panel (a). All points have been computed at a low noise amplitude of $\eta=0.2$.}
    \label{fig:parameters_volume}
\end{figure*}

When studding the system structure near the transition, all cases show a very similar behavior, showing a wide band traveling though a disordered phase as shown in Fig. \ref{fig:plane_beta}(a). This means that the effects of breaking the forward-backward symmetry are progressively reduced when the noise contribution increase. Despite the important differences between the systems at low noise values, Figs. \ref{fig:plane_beta}(c) and (d) ends showing the same values at medium and large noises.

In contrast to what was observed in the previous section when varying the aperture angle $\alpha$, the orientation angle $\beta$ does not significantly affect the global order-disorder transition. As illustrated in Fig. \ref{fig:parameters_beta}(c), the critical noise $\eta_c$ - obtained from the minimum of the Binder cumulant (Fig. \ref{fig:parameters_beta}(b)) - remains remarkably constant at $\eta_c \approx 0.55$ across all studied values of $\beta$. Despite the profound differences in the spatial structure of the systems (ranging from dense traveling bands to highly homogeneous states), the polar order parameter $\psi$ behaves almost identically in all cases (see Fig. \ref{fig:parameters_beta}(a)). This behavior provides strong support for the hypothesis that the global polar order as a function of noise depends solely on the total field-of-view area (which is constant here since $\alpha=90^\circ$ is fixed), rather than on the specific orientation or directionality of the vision cones.

To further characterize these states, Fig. \ref{fig:parameters_beta}(f) shows the parametric relationship between the global polar order $\Psi$ and density fluctuations $\sigma_\rho^2$. For backward-biased systems ($\beta = 135^\circ$), the trajectory describes a wide loop with a very high fluctuation peak at intermediate order, which corresponds to the coexistence phase near the transition. In contrast, for forward-biased systems ($\beta = 45^\circ$), this loop is almost completely flattened, typical of a strong first-order-like coexistence regime with prominent traveling bands, highlighting a direct and smooth transition from disorder to order with significant density clustering.

To complete this picture, Figs. \ref{fig:parameters_beta}(g)-(i) show the local neighborhood properties. At high noise, where the system is in a fully disordered gas phase \cite{Vicsek1995}, all curves for different values of $\beta$ converge to the same isotropic limit. In this random state, the number of neighbors inside the vision cone (Fig. \ref{fig:parameters_beta}(i)) is simply proportional to the global mean density and the geometric area of the cones ($N_v \approx \langle\rho\rangle \alpha R_0^2 \approx 4.71$).  However, at low noise (ordered phase), the spatial self-organization dynamically alters these values. The forward-biased configuration ($\beta = 45^\circ$) exhibits a slightly higher average number of isotropic neighbors $\langle N_0 \rangle$ than the backward-biased cases (Fig. \ref{fig:parameters_beta}(g)), which is consistent with appearance of empty regions and narrow bands. This huge number of neighbors is one of the reasons that motivate us to study the volume exclusion effect in this system.

\subsection{Volume exclusion effect}
\label{sec:exclusion}

In this final section, we investigate the effects of introducing volume exclusion by varying the ratio between the particle diameter and the interaction radius, $\sigma/R_0$. Our ultimate goal is to bridge the gap between theoretical models and actual experimental realizations, providing a better understanding of how the previously identified non-reciprocal phases survive under physical space constraints.

To achieve this, we study four distinct representative cases at a constant, relatively low noise level of $\eta=0.20$. We classify these cases based on the spatial structures they present in the point-particle limit. We compare the classic Vicsek Model ($\alpha=180^\circ, \beta=90^\circ$) with the symmetric lateral vision case ($\alpha=90^\circ, \beta=90^\circ$). Both models exhibit uniform flocking in the absence of volume. This comparison aims to determine if steric interactions trigger any experimentally observable micro-structural differences in an otherwise homogeneous phase. We examine two distinct mechanisms of band formation: the narrow-vision bands ($\alpha=10^\circ, \beta=90^\circ$) and the front-biased bands ($\alpha=90^\circ, \beta=45^\circ$). This allows us to understand how steric repulsions disrupt high-density structures formed by different symmetrical breakings. Based on the snapshots of Fig. \ref{fig:parameters_volume}(a), there is no visible difference between the first couple of cases, but there is an important contrast with the banded ones; while the narrow-vision bands are not observed, the front-biased band maintain their band structure for a reasonable $\sigma/R_0$ ratio and a small number of particles, which make them perfect to observe in an experimental system.\\

As shown in the structural snapshots of Fig. \ref{fig:parameters_volume}(a), steric interactions severely limit the maximum local packing fraction of the system. In the banded phases, which inherently rely on particles overlapping or packing extremely close to one another to maintain the dense traveling wave, the introduction of a finite size $\sigma$ acts as an effective disruptive force. As the particle size increases, these ultra-dense forward-biased bands expand, lose their sharp boundaries, and eventually ''melt'' into more uniform configurations. This melted phase is presented in all of our cases for a ratio of $0.40$ which, for a number density of $3.0$ means it almost fills every space inside the box. Despite this, Fig. \ref{fig:parameters_volume}(b) shows that the order parameter $\Psi$ is not significantly affected by $\sigma$ in any of the cases studied, which suggest that volume exclusion only has a role in the structure formation, but not in the polar order of the system, as is already known for the isotropic Vicsek Model \cite{martinez2018collective}. Physically, this occurs because the short-range repulsive forces maintain neighbors away enough to avoid them being in the same place but they still are inside the interaction distance $R_0$. The addition of excluded volume acts as a structural noise contribution, which does not affect the orientation of particles, but evolves every system into an homogeneous liquid.


\section{Conclusions}
\label{sec:conclusions}

In this work, we have introduced and extensively characterized a modified Vicsek Model that incorporates biologically inspired blind spots through two distinct, non-overlapping vision cones. By systematically varying the angular aperture ($\alpha$) and the orientation ($\beta$) of the cones, we demonstrated that spatial anisotropy and non-reciprocity act as powerful microscopic mechanisms for pattern formation in active systems.

Our findings indicate that restricting the lateral field of view (decreasing $\alpha$) lowers the critical noise required for the order-disorder transition and heavily enhances local clustering. This restriction leads to the formation of ultra-dense, narrow traveling bands, and uniquely triggers the emergence of a perpendicular ''cross sea'' phase, highlighting a shift towards a highly discontinuous phase separation.

Moreover, manipulating the front-back symmetry (varying $\beta$) unraveled the drastic effects of non-reciprocal interactions. Forward-biased vision induces a ''follow-the-leader'' pursuit mechanism that destabilizes homogeneous flocks into highly dense parallel bands. In stark contrast, backward-biased vision acts as a negative feedback loop with even more long-range correlation than the symmetrical cases, which are the ones that have the least density fluctuation for low noises. Strikingly, while the spatial morphology is highly sensitive to the vision directionality, the onset of global polar order remains invariant as long as the total vision area is conserved. 

Finally, the introduction of short-range steric interactions (volume exclusion) revealed that the ultra-dense banded phases are highly sensitive to the agents' physical dimensions, but perfectly replicable in experiments with a front-biased interaction system. Steric collisions effectively inject internal structural noise into the system,  melting the tightly packed non-reciprocal structures into more uniform configuration while maintaining the orientational alignment.

Looking forward, several promising avenues for future work emerge from this study. First, different explorations in the parameter space could be done, including breaking other symmetries of the interaction matrix, such as the left-right symmetry.  Second, deriving a coarse-grained continuum hydrodynamic theory for this dual-cone system could provide deeper analytical insights into the stability of the phases presented in the system. Ultimately, our findings open the door to direct experimental realizations, particularly in robotic swarms or active granular matter, where both programmable restricted perception and volume exclusion naturally coexist.

\section{Funding and Acknowledgments}

C. V. acknowledges fundings IHRC22/00002 and Proyecto PID2022-140407NB-C21 funded by MCIN/AEI /10.13039/501100011033 and FEDER, UE. J. M. R. acknowledges funding from Juan de la Cierva postdoctoral fellowship funding from Spanish Ministry of Education (JDC2024-053228-I).

\section{Author contributions}

All authors contributed to the conception of the work. They also were involved in drafting or revising the article critically for intellectual content. All approved the final version.

\section{Data Availability Statement }

The data that supports the findings of this study are available from the corresponding author upon reasonable request.

\section{Bibliography}


%

\end{document}